\documentstyle[aps,12pt]{revtex} 
\begin{document}
\renewcommand{\thefootnote}{\fnsymbol{footnote}}

\author{ Youshan Dai $^{a,b}$ and Dongsheng Du $^{a}$ \\ 
{\small $a$.  Institute of High Energy Physics, Chinese Academy of Sciences }\\ {\small P.O.Box 918(4),
Beijing, 100039, P.R.China }\\ 
{\small $b$.  Department of Physics, Hangzhou University, Hangzhou, 310028, 
P.R.China }}

\title{ 
{\Large\sf CP Violation in Two-Body Hadronic Decays of $B_c$ Meson }
\footnote{Supported in part by National Natural Science Foundation of
China.} } 
\maketitle

\thispagestyle{empty} 
\begin{abstract}
 
 Using the next-to-leading order low energy effective Hamiltonian, the CP
asymmetries for the $B_{c}$ meson decays into meson pair are calculated in the
spectator approximation.  We do not compute the hadronic matrix elements
directly, instead, we use the amplitude ratios to estimate the CP asymmetries.
This is quite,different from the previous works in the literature.
   The values of the momentum squared carried by the virtual particles
in timelike penguin dirgrams are also carefully discussed. From our calculated
results, the best decay modes to observe CP violation in $B_{c}$ decays would be
$B_{c}^-\rightarrow{\bar D}^{*0}K^{*-}$ , ${\bar D}^{0}K^{*-}$ , ${\bar D}^{*0}
K^{-}$ , ${\bar D}^{0}K^{-}$ and $B_{c}^-\rightarrow{\eta}_{c}D^{-}$, which 
need about $10^8$ of $B_{c}^{\pm}$ events in experiment.

\end{abstract}

\newpage 
\section*{I.  Introduction} 
\noindent

  One of the main aims of B factories  is the observation of CP
violation.  The $B_{u}, B_{d}$ and $B_{s}$ meson decays and CP violation have
been discussed extensively.  The decays of $B_{c}$ meson ($\bar{b}c$ and b$\bar
{c}$ bound states) seems to be another valuable window for probing the origin of
CP violation.  Since large number of $B_{c}$ mesons is expected to be producted
at hadronic colliders like LHC or Tevatron [1], to examine the features of the
$B_{c}$ meson decays and CP violation becomes more and more intersting for both
experimental efforts and theoretical studies.

  Theoretical predictions about $B_{c}$ meson decays are made in many previous
works with different models [2].  The  results are strongly
model-dependent.  It is difficult to judge these different results for lack of
experimental data for the time being.  There are also several works which pay
attention to  CP violation in $B_{c}$ meson decays.  In Ref.[3] the BSW mode
and in Ref.[4] the Bethe-Salpeter formulism are used to calculate CP
violating asymmetries for two-body mesonic decays of the $B_{c}$ meson.  There
are many  uncertainties due to the model-dependence and different
choice of parameter values .

  It is well known that there is only direct CP violation in $B_{c}$ meson
decays which requires that two decay amplitudes must have both different weak
phases and different strong phases.  The weak phases come from the CKM matrix.
According to the new experimental results, the favorite values of the
Wolfenstein parameters for the CKM matrix are changed, the main difference
compared with the old ones is the sign of the parameter $\rho$ [5].
Obviously, the new Wolfenstein
parameter will influence the weak phases and the calculated results of the
CP violating asymmetry. If we ignore  soft strong phases which are
difficult to estimate at present,  hard strong phases can be estimated
perturbatively
using the low effective Hamiltonian and are relevant to the loop-integral
functions of penguin diagrams.  In general, decay amplitudes and the CP
violating asymmetries rely on hadronic matrix elements which are
strongly
model-dependent, but in the spectator approximation (i.e.  ignore the
contributions of the annihilation and spacelike penguin diagrams, which
are believed to
be form factor and color suppressed),  CP violating asymmetries can be
obtained without calculation of the hadronic matrix elements for many decay
processes of the $B_{c}$ meson.  In this paper, unlike the previous works in the
literature [3,4], we try to estimate the CP violating asymmetry of the
$B_{c}$ decays
into meson pair without calculating the hadronic matrix elements directly.  
So that part of the uncertainties caused by the direct computation of the
hadronic matrix elements are avoided. Of course we still need to use 
factorizition approximation to factorized out some coefficients of the
decay amplitudes. Thus the factorization approximation will
definitely cause uncertainties.  Another uncertainty source is the unknown
momentum squared carried by the virtual particles in penguin diagrams.  In the
literature, one used to pick up a special value of $k^2$ from [0 , $m_{b}^2$] or
[$\frac{1}{4}m_{b}^2$ , $\frac{1}{2}m_{b} ^2$] for all timelike penguin
daigrams.  Obviously, it is a rather rough approximation.  We shall discuss this
problem in our paper and try to avoid the drawback of taking arbitrary
values for
$k^2$ based on the valence-quark assumption.  In this article, we use the
next-to-leading order low energy effective Hamiltonian [6] to calculate the CP
violating asymmetry for the $B_{c}$ decays into PP, PV and VV mesons in the
spectator approximation (P- pseudoscalar meson, V- vector meson).  In
Section II, we first list the  formula for the CP violating
asymmetry.  The quark-diagram amplitudes using the next-to-leading order low
energy effective Hamiltonian are given in Section III.  In Section IV, we
discuss the value of $k^2$ carried by the virtual particles for the timelike
penguin dirgrams.  The calculated results of the CP violating asymmetries for
the $B_c$ decays into meson pair are given in Section V.  Finally 
Section VI is devoted to discussion and conclusion.

\section*{II.  Formula of CP Asymmetry} 
\noindent

  For the $B_c$ meson decays to a final state $f$, we may, without loss of
generality, write the decay transition amplitude as $$A(f)=G_1T_1 + G_2T_2 \eqno
(2.1) $$ where $G_1$, $G_2$  are both multiplication of the CKM matrix
elements. Assuming CPT invariance, the CP-conjugated amplitude is
$$
\bar{A}(\bar{f})=G_1^*T_1 + G_2^*T_2 
\eqno (2.2)
$$ 
With the help of (2.1) and(2.2), the CP violating asymmetry can be written as 
$$ 
{\cal A}_{cp}(f)\equiv\frac{|A(f)|^2-|\bar{A}(\bar{f})|^2}{|A(f)|^2+|\bar{A}
(\bar{f})|^2}=\frac{-2(ImG_1G_2^*)(ImT_1T_2^*)}{|G_1T_1|^2+|G_2T_2|^2+
2(ReG_1G_2^*)(ReT_1T_2^*)} 
\eqno (2.3) 
$$ 
Define $G_1=|G_1|e^{i\theta_1}$, $G_2=|G_2|e^{i\theta_2}$ and $T_1=|T_1|e^ 
{i\delta_1}$, $T_2=|T_2|e^{i\delta_2}$, where $\theta_1$, $\theta_2$ are the 
weak phases and $\delta_1$, $\delta_2$ are the strong phases respectively, then 
we have 
$$
{\cal A}_{cp}(f) =~\frac{-2sin({\theta_1}-{\theta_2})sin({\delta_1}-{\delta_2})}
{|\frac{G_1T_1}{G_2T_2}|+|\frac{G_2T_2}{G_1T_1}|+2cos({\theta_1}-{\theta_2})
cos({\delta_1}-{\delta_2})} 
\eqno(2.4) 
$$ 
Obviosly, the decay amplitude must have both different weak phases 
(${\theta_1}\not={\theta_2}$) coming from the CKM matrix and different strong 
phases (${\delta_1}\not={\delta_2}$) coming from final state interactions (FSI),
otherwise there are no CP violation. Introduce another angle $\zeta$ which is 
defined as 
$$
h\equiv\left | \frac{G_1T_1}{G_2T_2}\right | \equiv
tg\frac{\zeta}{2}~~, ~~~~~~~sin{\zeta}=\frac{2h}{1+h^2}>0 
\eqno(2.5) 
$$
then the CP violating asymmetry formula becomes 
$$ 
\begin{array}{l} {\cal A}_{cp}(f)=\frac{{\cal A}_{0}}{1+{\cal A}_{1}}\\ 
{\cal A}_{0}=-sin{\zeta}sin({\theta_1}-{\theta_2})sin({\delta_1}-{\delta_2})\\ 
{\cal A}_{1}=sin{\zeta}cos({\theta_1}-{\theta_2})cos({\delta_1}-{\delta_2})
\end{array} 
\eqno(2.6) 
$$ 
We can see that the magnitude of the direct CP violation depends on three 
angles:  weak phase ($\theta_1$-$\theta_2$), strong phase ($\delta_1$- 
$\delta_2$) and the angle $\zeta$.

(i)The weak phase ($\theta_1$-$\theta_2$) is decided by the CKM matrix
elements, sin($\theta_1$-$\theta_2$)=$\frac{ImG_1G_2^*}{|G_1G_2|}$, cos(
$\theta_1$-$\theta_2$)=$\frac{ReG_1G_2^*}{|G_1G_2|}$.  According to the
Wolfenstein representation of the CKM matrix [7] 
$$ 
V=\left( \begin{array}{lcr}
V_{ud} & V_{us} & V_{ub}\\ V_{cd} & V_{cs} & V_{cb}\\ V_{tb} & V_{ts} & V_{tb}
\end{array} \right)=\left( \begin{array}{ccc} 1-\frac{\lambda^2}{2} & \lambda &
A\lambda^{3}(\rho-i\eta)\\ -\lambda & 1-\frac{\lambda^2}{2} & A\lambda^{2}\\
A\lambda^{3}(1-\rho-i\eta) & -A\lambda^{2} & 1 
\end{array}
\right)+O(\lambda^{4}) 
\eqno (2.7) 
$$ 
For two-body mesonic $B_c^-$ decays, if we only consider $ b$ quark
decay, we have, 
$|G_1G_2|=|V_{ud}^*V_{cd}^*V_{ub}V_{cb}|=|V_{us}^*V_{cs}^*V_{ub}V_{cb}|
\approx A^2\lambda^6(1-\frac{\lambda^2}{2})\sqrt{\rho^2+\eta^2}$, and $ImG_1G_2
^*=\pm J \approx\pm A^2\lambda^6(1-\frac{\lambda^2}{2})\eta$.  It followes that
$$sin(\theta_1-\theta_2)=\pm \frac{\eta}{\sqrt{\rho^2+\eta^2}}~~,~~~~|\theta_1-
\theta_2|=\gamma 
\eqno(2.8) 
$$ 
where the angle $\gamma$ is just one of the angles of the unitarity triangle 
for the CKM matrix elements in Fig.1.  From (2.8) , we see that the weak phase 
($\theta_1-\theta_2$) is only decided by the Wolfenstein parameters 
($\rho~,~\eta$).

(ii)The strong phase ($\delta_1-\delta_2$) is caused by final state
interactions, in which the penguin effects (hard strong phases) can be
estimated
perturbatively and the rescattering effects (soft strong phases) is
unknown at present.  In this paper we will ignore the rescattering effects
and disscuss the penguin effects only.

(iii)The angle $\zeta$ relies on the ratio $h=|\frac{G_1T_1}{G_2T_2}|=tg\frac
{\zeta}{2}$.  Although the decay amplitudes $T_1$ and $T_2$ both can be
calculated, but there are uncertainties since the calculation of hadronic
matrix elements, therefore$T_1, T_2 $, are modle dependent.  However, for
many $B_c$ decay
processes, we do not need to compute the decay amplitudes (hadronic matrix
elements) individually because the ratio $\frac{T_1}{T_2}$ is
independent of the hadronic matrix elements ( in factorization
approximation). For example, in our formula of the CP
violating asymmetry, it only needs to focus on calculating the ratio $h$ and
$sin\zeta =f(h)=\frac {2h}{1+h^2}$.   In the case of $h\ll 1(or~ h\gg 1)$, 
we have $sin\zeta =f(h)\approx
2h(or~\frac{2}{h})\ll 1,~ |{\cal A}_{1}|\ll$1, so the CP violating asymmetry
(2.6) can be simplified as 
$$ 
{\cal A}_{cp}(f)=\frac{{\cal A}_{0}}{1+{\cal
A}_{1}}\approx {\cal A}_{0}= -sin\zeta
sin(\theta_1-\theta_2)sin(\delta_1-\delta_2)~~~(for~~h\ll 1~~or~~h\gg 1)
\eqno(2.9) 
$$

\section*{III.  The Quark-Diagram Amplitudes} 
\noindent

  Following Ref.[6], the next-to-leading order low energy effective Hamiltonian
for $|\Delta B|$=1 is given at the renomalization scale $\mu \sim m_{b}$ as
 
$$
{\cal H}_{eff}(|\Delta B|=1)=\frac{G_F}{\sqrt{2}}\sum\limits_{\stackrel
{q=u,c}{q'=d,s}}V_{qq'} V_{qb}^{*}\left\{ C_{1}(\mu )Q_{1}+C_{2}(\mu )Q_{2}+
\sum\limits_{k=3}^{10} C_{k}(\mu )Q_{k}\right\}~~+~~h.c.  \eqno(3.1) $$
 
where the Wilson coefficients  $C_{i}(\mu)~(i=1,2,\cdots,10)$ are 
calculated in the renormalization group improved pertubation theory and include 
the leading and next-to-leading order QCD corrections.  $Q_1$ and $Q_2$ are the 
tree diagram operators; $Q_{3},\cdots,Q_{6}$ are the QCD penguin diagram 
operators; whereas $Q_{7},\cdots,Q_{10}$ are the electroweak penguin diagram 
operators.  
$$
\begin{array}{ll}
Q_{1}=(\bar{q'}_{\alpha}q_{\beta})_{V-A}(\bar{q}_{\beta}b_{\alpha})_{V-A} &
Q_{2}=(\bar{q'}q)_{V-A}(\bar{q}b)_{V-A}\\
Q_{3(5)}=(\bar{q'}b)_{V-A}\displaystyle\sum_{q"}(\bar{q"}q")_{V-A(V+A)} &
Q_{4(6)}=(\bar{q'}_{\alpha}b_{\beta})_{V-A}\displaystyle\sum_{q"}(\bar{q"}_
{\beta}q"_{\alpha})_{V-A(V+A)}\\
Q_{7(9)}=\frac{3}{2}(\bar{q'}b)_{V-A}\displaystyle\sum_{q"}e_{q"}
(\bar{q"}q")_{V+A(V-A)} &
Q_{8(10)}=\frac{3}{2}(\bar{q'}_{\alpha}b_{\beta})_{V-A}\displaystyle\sum_{q"}
e_{q"}(\bar{q"}_{\beta}q"_{\alpha})_{V+A(V-A)} 
\end{array} 
\eqno(3.2) 
$$ 
In (3.2), $q"$ is runing over the quark flavors being active at the scale 
$\mu\sim m_{b}~(q"\in u,d,s,c,b)$, $e_{q"}$ are the corresponding quark charges
in unite of $|e|$, $\alpha$ and $\beta$ are $SU(3)_{c}$ color indices, $(V\pm A)$ 
refer to $\gamma_{\mu}(1\pm\gamma_{5})$.  In general, the nonleptonic $B_c$ 
decays can occur through both the spectator channels and nonspetator
channels. The latter are difficult to deal with at present and are commonly 
assumed to be form factor suppressed compared to the former ones.  
In spectator approximation , the two-boty mesonic $B_c^-$ decay can be 
described by quark diagrams as in 
Fig.2. Using ${\cal H}_{eff}$ (3.1) in a renormalization-scheme independent 
way [8] and factorization approximation, the quark-diagram amplitudes of
Fig.2.(where the CKM matrix elements have
been singled out) can be obtained 
$$
\begin{array}{ll} T(a^{tree})=a_{1}A~~~~&
T(a_{q}^{pen})=[(a_{3}+a_{9})+\xi_{f}(a_{5}+a_{7})+(1+\xi_{f})b_1G_{q}(k^{2})]A\\
T(b^{tree})=a_{2}B &
T(b_{q}^{pen})=[(a_{3}-\frac{1}{2}a_{9})+\xi_{f}(a_{5}-\frac{1}{2}a_{7})+
(1+\xi_{f})b_2G_{q}(k^{2})]B 
\end{array} 
\eqno(3.3) 
$$ 
where $T(a^{tree})$ refer to the amplitude of the diagram $a^{tree}$ in Fig.2. 
etc., A and B denote the factorized hadronic matrix elements 
$$ 
\begin{array}{ll} A\equiv\frac{G}{\sqrt
2}<X^{-}|(\bar{q}_{1}q_{2})_{V-A}|0><X^{0}|(\bar{q}_{3}b) _{V-A}|B_c^- >~~~~&
for X^- =(q_1\bar{q}_2 )~~,~~X^0 =(q_3\bar{c})\\ B\equiv\frac{G}{\sqrt
2}<X^{0}|(\bar{q}_{1}q_{2})_{V-A}|0><X^{-}|(\bar{q}_{3}b) _{V-A}|B_c^- >~~~~&
for X^0 =(q_1\bar{q}_2 )~~,~~X^- =(q_3\bar{c}) 
\end{array} 
\eqno(3.4) 
$$ 
In (3.3), $\xi_{f}$ arises from the transformation of (V-A)(V+A) currents into
(S+P)(S-P) and further into (V-A)(V-A) ones for $Q_{5-8}$ 
$$
\xi_{f}=\left\{\begin{array}{ll}
\frac{(1+\eta_{X^-})M_{X^-}^{2}}{(m_{q'}+m_{q_v})(\eta_{X^0}m_{b}-m_{q_v})}
~~~~& for~a^{pen}~diagrams\\
\frac{(1+\eta_{X^0})M_{X^0}^{2}}{(m_{q'}+m_{q'_v})(\eta_{X^-}m_{b}-m_{q'_v})}
~~~~& for~b^{pen}~diagrams 
\end{array} \right.  
\eqno(3.5) 
$$ 
where 
$$
\eta_{X^0}(or~\eta_{X^-})=\left\{\begin{array}{ll} +1~~~~& for~ X^0 (or~ X^-
)=P~(pseudoscalar~meson)\\ -1~~~~& for~ X^0 (or~ X^- )=V~(vector~meson)
\end{array} \right.  
\eqno(3.6) 
$$ 
From (3.5) and (3.6), we can see that $\xi_{f}=0$ when $X^-$ is vector meson in 
$a^{pen}$ diagrams and when $X^0$ is vector meson in $b^{pen}$ diagrams.  The 
reason is the following: using the Fierz rearrangements for
$Q_{5-8}$, the calculated results of the decay amplitudes are 
proportional to the factor
$P^{\mu}(X^{-})\epsilon_{\mu}(X^{-})(or~P^{\mu}(X^{0})\epsilon_{\mu}(X^{0}))$
which is equal to zero ($P^\mu $ and $\epsilon_\mu $ are momentum and
polarization vectors of the vector meson respectively).  In (3.3), $a_k$
is
defined as 
$$
a_{2i-1}\equiv\frac{\bar{c}_{2i-1}}{N_c}+\bar{c}_{2i}~~~~a_{2i}\equiv\frac
{\bar{c}_{2i}}{N_c}+\bar{c}_{2i-1}~~~~(i=1,2,3,4,5) 
\eqno (3.7) 
$$ 
where $\bar{c}_{k}~(k=1,\cdots,10)$ are the renormolization scheme independent 
Wilson coefficients $\bar{c}_{k}(\mu)$ at $\mu\sim m_{b}$.  We use $\alpha_{s}
(m_{Z})=0.118,~ \alpha_{em}(m_{Z})=\frac{1}{128},~ m_{t}=174 GeV,~\bar{c}_{i}$
is taken as [9] 
$$ 
\begin{array}{llll} 
\bar{c}_{1}=-0.313 & \bar{c}_{2}=1.150 &~ & ~ \\ 
\bar{c}_{3}=0.017 & \bar{c}_{4}=-0.037 & \bar{c}_{5}=0.010 &\bar{c}_{6}=-0.047 \\
\bar{c}_{7}=-0.001\alpha_{em}& \bar{c}_{8}=0.049\alpha_{em} & \bar{c}_{9}=-1.321 \alpha_{em} &
\bar{c}_{10}=0.267\alpha_{em} 
\end{array} 
\eqno(3.8) 
$$ 
and 
$$ 
\begin{array}{l}
b_1=\frac{2}{3}\left[\frac{\alpha_s}{8\pi}\bar{c}_{2}(1-\frac{1}{3N_{c}})+
\frac{\alpha_{em}}{3\pi}(\bar{c}_{1}+\frac{\bar{c}_2}{3})\frac{1}{N_c}\right]\\
b_2=\frac{2}{3}\left[\frac{\alpha_s}{8\pi}\bar{c}_{2}(1-\frac{1}{3N_{c}})-
\frac{\alpha_{em}}{6\pi}(\bar{c}_{1}+\frac{\bar{c}_2}{3})\frac{1}{N_c}\right]
\end{array} 
\eqno(3.9) 
$$ 
$$
G_{q}(k^{2})=\frac{3}{2}\left[\frac{10}{9}-F_{q}(k^{2})\right] 
\eqno(3.10) 
$$
where $F_{q}(k^{2})$ denotes the penguin loop-integral function with momentum
transfer squared $k^2$ at the scale $\mu\sim m_{b}$ 
$$
F_{q}(k^{2})=-4\int\limits_{0}^{1}dxx(1-x)ln\left[\frac{m_{q}^{2}-x(1-x)k^{2}}
{m_b^2}\right] 
\eqno(3.11) 
$$ 
Supposing $k^{2}>4m_{q}^{2}~~(q=u,c)$, the parameter $r_{q}=\frac{4m_q^2}{k^2}
<1$, the function $G_{q}(k^{2})$ can be analytically expressed as [10] 
$$
G_{q}(k^{2})=ln\frac{m_q^2}{m_b^2}-r_{q}+(1+\frac{r_q}{2})\sqrt{1-r_{q}}ln\frac
{1+\sqrt{1-r_{q}}}{1-\sqrt{1-r_{q}}}+i\pi (1+\frac{r_q}{2})\sqrt{1-r_{q}}
\eqno(3.12) 
$$ 
In the case of $q=u$, since $r_{u}\ll 1$, we have $G_{u}(k^{2})\approx ln\frac 
{k^2}{m_b^2}+i\pi$, the absorptive part of the $G_{q}(k^{2})$ will lead to  
direct CP violation.

\section* {IV.  Momentum Squared Carried by The Virtual Particles} 
\noindent

  The penguin loop-integral function $F_{q}(k^{2})$ depends crucially on the
$k^2$ carried by the virtual gluon, photon and $Z^0$.  In the literature, one
used to pick up a special fixed value of $k^2$ for timelike penguin diagrams
from [0 , $m_b^2$] or [$\frac{1}{4}m_{b}^{2}$ , $\frac{1}{2}m_{b}^{2}$] [10], 
but in general, taking the same value of $k^2$ for all different quark
diagrams is
not correct.  This problem has been discussed in Ref.[11], we shall
further discuss
it carefully in the following.  For timelike penguin diagram of the two-body
mesonic decay $B_c^- \rightarrow$ XY as illustrated in Fig.3, using the
4-momentum conservation ($\bar c$ as spectator quark), we have 
$$ 
\left\{ \begin{array}{lll} k^{2} &=&
m_{b}^{2}+m_{q}^{2}-2E_{b}E_{q}+2\stackrel{\rightarrow}{p_b}\cdot
\stackrel{\rightarrow}{p_q}\\ \stackrel{\rightarrow}{p_b} &=&
\stackrel{\rightarrow}{p_q} +\stackrel{\rightarrow} {p}_{\bar{q}_{v}}
+\stackrel{\rightarrow}{p}_{q_v}\\ E_{b} &=& E_{q}+E_{\bar{q}_{v}}+E_{q_v}
\end{array} \right.  
\eqno(4.1) 
$$ 
In the rest frame of the X meson, $\stackrel{\rightarrow}{p_X}=\stackrel
{\rightarrow}{p_q}+\stackrel{\rightarrow}{p_{\bar{q}_v}}=0$.  Denote
$a\equiv
E_{q}+E_{\bar{q}_{v}}$, from (4.1), wew have 
$$ 
\left\{ \begin{array}{rll}
|\stackrel{\rightarrow}{p_b}| &=& |\stackrel{\rightarrow}{p}_{q_v}|\\ E_b &=&
a+E_{q_v} \end{array} \right.~~~~~~~~\left\{ \begin{array}{rll}
|\stackrel{\rightarrow}{p_q}| &=& |\stackrel{\rightarrow}{p}_{\bar{q}_{v}}|\\
E_q &=& a-E_{\bar{q}_{v}} 
\end{array} \right.  
\eqno(4.2) 
$$ 
then we obtain 
$$ \begin{array}{ll} 
E_{b}=\frac{1}{2a}({m_b^2}+{a^2}-{m_{q_v}^2})~~,~~& |\stackrel{\rightarrow}{p_b}|
=\frac{1}{2a}\sqrt{[(m_{b}+a)^{2}-{m_{q_v}^2}][({m_b}-a)^{2}-{m_{q_v}^2}]}\\
E_{q}=\frac{1}{2a}({m_q^2}+{a^2}-{m_{q_v}^2})~~,~~&
|\stackrel{\rightarrow}{p_q}|
=\frac{1}{2a}\sqrt{[(m_{q}+a)^{2}-{m_{q_v}^2}][({m_q}-a)^{2}-{m_{q_v}^2}]}
\end{array} 
\eqno(4.3) 
$$ 
Because $q,~\bar{q}_{v}$ form a bound state X, in the valence-quark assumption, 
there are
${p_q}={\eta_q}{p_{X}}+p~,~{p_{\bar{q}_{v}}}=\eta_{\bar{q}_{v}}{p_{X}} -p$ ,
where $\eta_{q}=\frac{m_q}{m_{q}+m_{q_v}}~,~\eta_{\bar{q}_{v}}=\frac
{m_{q_v}}{m_{q}+m_{q_v}}$ ; $p_X$ is the total 4-momentum of the bound state X
and $p$ is the relative 4-momentum of $q$ vs.  $\bar{q}_v$ inside the X meson.
In the rest frame of the X meson, it is easy to get ${E_q}=\frac{1}{2M_X}
({m_q^2}+{M_X^2}-{m_{q_v}^2})$.  So we have $a={M_X}$.  In eq.(4.1), taking
$\stackrel{\rightarrow}{p_b}\cdot\stackrel{\rightarrow}{p_q}=|\stackrel
{\rightarrow}{p_b}||\stackrel{\rightarrow}{p_q}|cos\varphi$, where $\varphi$ is
the angle between $\stackrel{\rightarrow}{p_b}$ and
$\stackrel{\rightarrow}{p_q}$ in the rest frame of the X meson, $\varphi
=\angle(\stackrel{\rightarrow}{p_b}~,~ \stackrel{\rightarrow}{p_q})$, then we
get 
$$
k^{2}(\varphi)={m_b^2}+{m_q^2}-2{E_b}{E_q}+2|\stackrel{\rightarrow}{p_b}|\stackrel
{\rightarrow}{p_q}|cos\varphi~~,~~ \bar{k^2}=\frac{1}{\pi}\int\limits_{0}^{\pi}
d\varphi{k^2}({\varphi})={m_b^2}+{m_q^2}-2{E_b}{E_q} 
\eqno(4.4) 
$$ 
$$
({k^2})_{max}={m_b^2}+{m_q^2}-2{E_b}{E_q}+2|\stackrel{\rightarrow}{p_b}||\stackrel
{\rightarrow}{p_q}|~~,~~({k^2})_{min}={m_b^2}+{m_q^2}-2{E_b}{E_q}-|\stackrel
{\rightarrow}{p_b}||\stackrel{\rightarrow}{p_q}| 
\eqno(4.5) 
$$ 
where $E_b$, $E_q$, $|\stackrel{\rightarrow}{p_b}|$, $|\stackrel{\rightarrow} 
{p_q}|$ can be obtained from eq.(4.3) by taking $a=M_X$. Accordingly we
find that the average value of the $k^2$ is 
$$ 
\frac{\bar{k^2}}{m_b^2}=\frac{1}{2}\left[1+\frac{{m_{q_v}^2}-{m_q^2}}{M_X^2}
(1-\frac{m_{q_v}^2}{m_b^2})+\frac{{m_q^2}+2{m_{q_v}^2}-{M_X^2}}{m_b^2}\right]
\eqno(4.6) 
$$ 
For the timelike penguin diagram,
$0\leq({k^2})_{min}\leq{k^2}(\varphi)\leq({k^2}) _{max}\leq{m_b^2}$.  Since
$\varphi$ is unknown in experiment, we take the average value of the $\bar{k}^2$
to calculate the penguin loop-integral function.  Taking the quark masses
$(m_u,~m_d,~m_s,~m_c,~m_b)=(0.005,~0.01,~0.175,~1.35,~4.8)$ GeV and using the
meson masses $M_X$ according to the particle data [12], the values
$\frac{\bar{k^2}}{m_b^2}$ of the timelike penguin diagram for different $B_c^-$
decay processes are given in Table II.

\section* {V.  CP Asymmetry} 
\noindent

  As discussed in Section II, the CP violation asymmetry of $B_c$ meson
decays into meson pair depends on the three angles:  weak phase
$({\theta_1}-{\theta_2})$, strong phase $({\delta_1}-{\delta_2})$ and 
the angle $\zeta$.  The weak phase $({\theta_1}-{\theta_2})$ is 
determined by the Wlfenstein parameter
$(\rho~,~\eta)$, the strong phase $({\delta_1}-{\delta_2})$ and the 
angle$\zeta$ are decided by the quark-diagram amplitudes.  For many decay 
processes, there is
no need to calculate the hadronic matrix elements because in the
factorization approximation, the factorized quark-diagram amplitudes 
depend only on a
single (dominant) hadronic matrix element (A or B) which is cancelled in the
ratio $\frac{T_1}{T_2}$.

  In below we discuss the decay process ${B_c^-}\rightarrow\bar{D^0}{\pi^-}$ as
an example to illustrate how to calculate the CP violating asymmetry in
our method. In the
spectator approximation, the decay amplitude for the ${B_c^-}\rightarrow
\bar{D^0}{\pi^-}$ is 
$$ 
A(f)=G_1T_1+G_2T_2={V_{ud}^*}{V_{ub}}[~T(a^{tree})+T(a_u^{pen})~]+
{V_{cd}^*}{V_{cb}}T(a_c^{pen}) 
\eqno(5.1) 
$$ 
using the quark-diagram amplitude calculated in section.III, we get 
$$
\frac{T_1}{T_2}=\left|\frac{T_1}{T_2}\right|e^{i(\delta_{1}-\delta_{2})}=
\frac{a_1+a_3+a_9+\xi_{({\bar{D}^0}{\pi^-})}(a_5+a_7)+[1+\xi_{({\bar{D}^0}
{\pi^-})}]b_1G_{u}(\bar{k^2})} {a_3+a_9+\xi_{({\bar{D}^0}{\pi^-})}(a_5+a_7)+
[1+\xi_{({\bar{D}^0}{\pi^-})}]b_1G_{c}(\bar{k^2})} 
\eqno(5.2) 
$$ 
$$ 
\left\{
\begin{array}{l} 
sin(\delta_{1}-\delta_{2})=\frac{1}{|T_1T_2|}(ImT_1 ReT_2-ImT_2ReT_1)\\ 
cos(\delta_{1}-\delta_{2})=\frac{1}{|T_1T_2|}(ReT_1 ReT_2+ImT_1 ImT_2)
\end{array} \right.  
\eqno(5.3) 
$$ 
where $\xi_{({\bar{D}^0}{\pi^-})}=\frac{2m_{\pi^-}^2}{(m_d+m_u)(m_b-m_u)} 
=0.545$, for $B_c^-\rightarrow{\bar{D^0}}{\pi^-}$ , $\frac{\bar{k^2}}
{m_b^2}=0.498~,~G_u(\bar{k^2})=-0.697+i\pi~,~G_c(\bar{k^2})=-2.059+i2.500$.
Taking $N_c=N_{c}^{eff}=2$, which is favored by experimental data [13], then the
numerical results can be obtained as $\left|\frac{T_1}{T_2}\right|=14.55~,~
sin(\delta_{1}-\delta_{2})=-0.197~,~cos(\delta_{1}-\delta_{2})=-0.980$ .  From
the Wolfenstein representation of the CKM matrix,  $G_1={V_{ud}^*}
V_{ub}=A\lambda^3(1-\frac{\lambda^2}{2})(\rho-i\eta)~,~G_2={V_{cd}^*}V_{cb}=
-A\lambda^3$ , it follows that
$$ 
\left\{ 
\begin{array}{l}
sin(\theta_1-\theta_2)=\frac{ImG_1G_2^*}{|G_1G_2|}=\frac{\eta}{\sqrt{\rho^2+
\eta^2}}\\
cos(\theta_1-\theta_2)=\frac{ReG_1G_2^*}{|G_1G_2|}=\frac{-\rho}{\sqrt{\rho^2+
\eta^2}} \end{array} \right.~~~~~~\left\{ \begin{array}{l}
h=\left|\frac{G_1T_1}{G_2T_2}\right|=\frac{\lambda^2\sqrt{\rho^2+\eta^2}}
{(1-\frac{\lambda^2}{2})}\left|\frac{T_1}{T_2}\right|\\ sin\zeta
=\frac{2h}{1+h^2} 
\end{array} \right.  
\eqno(5.4) 
$$ 
According to our CP asymmetry formula derived in section.II, we have ${\cal A}_
{cp} =\frac{{\cal A}_{0}}{1+{\cal A}_{1}}$, where ${\cal A}_{0}=-sin\zeta 
sin(\theta_1-\theta_2)sin(\delta_1-\delta_2)~,~{\cal A}_{1}=sin\zeta
cos(\theta_1-\theta_2)cos (\delta_1-\delta_2)$.  In this paper , we take
$\lambda=0.220~,~\eta=0.336~,~ \rho=0.160$ (the value of the Wolfenstein
parameter A has no effect on our calculation) [5].  For numerical
estimation, the results are $sin(\theta_1-\theta_2)=0.903~,~cos(\theta_1-\theta_2)
=-0.430~,~sin\zeta=0.365$ and ${\cal A}_{0}=0.0649~,~{\cal A}_{1}=0.154~,~
{\cal A}_{cp}(f=\bar{D^0}{\pi^-})=5.62 \times{10^{-2}}$.

  The quark-diagram amplitudes for the $B_c^-$ decays into PP, PV, and VV mesons
in the spectator approximation are given in Table.I.  It is seen that we
only need to compute the ratio $\frac{T_1}{T_2}$ and there is no need to
calculate the hadronic matrix elements directly in many cases, but for the
$B_c^-$ meson decays to $(c\bar{c})D^-~,~(c\bar{c})D^{*-}~,~(c\bar{c})D_s^-~
,~(c\bar{c})D_s^{*-}$ , the hadronic matrix elements A and B still appear in the
ratio 
$$
\frac{T_1}{T_2}=\frac{a_3+a_9+\xi_{f}(a_5+a_7)+(1+\xi_{f})b_1G_{u}(\bar{k^2})}
{a_1+a_2\frac{B}{A}+a_3+a_9+\xi_{f}(a_5+a_7)+(1+\xi_{f})b_1G_{c}(\bar{k^2})}
\eqno(5.5) 
$$ 
Fortunately $a_2$ is colour suppressed compared with $a_1$ and $\frac{B}{A} 
\sim O(1)$, so the calculated results are not sensitive to the
value of $\frac{B}{A}$ .  In this paper, we simply take $\frac{B}{A}=1$ for the
estimation of $CP$ asymmetries of these $B_c^-$ decay 
processes. 

  All calculated results of the CP violating asymmetries for two-body mesonic
decays of the $B_c^-$ meson are presented in Table II.

\section* {VI. Discussion and Conclusion}  
\noindent

  Using the next-to-leading order low energy effective Hamiltonian and the
quark-diagram amplitude method, we have calculated the CP violating asymmetries
for two-body mesonic decays of the $B_c$ meson.  For most decay processes
(except for $B_c^-\rightarrow (c\bar{c})D^-$ , $(c\bar{c})D^{*-}$ , $(c\bar{c})
D_s^-$ , $(c\bar{c})D_s^{*-}$ ), the CP violating asymmetries do not rely on
the hadronic matrix elements since they are canceled in the ratio
$\frac{T_1}{T_2}$.
So there is no model-dependence caused by calculating the hadronic matrix 
elements.

  The CP violating asymmetry is proportional to ${\cal A}_{0}=-sin\zeta sin
(\theta_1-\theta_2)sin(\delta_1-\delta_2)$.  From Table II, we see that
$sin(\theta_1-\theta_2)$ is the same for all decay processes (except for a sign)
which is  determined by the Wolfenstain parameters ($\rho$ , $\eta$) only.
In the processes (1)-(16),$sin(\theta_1-\theta_2)=\frac{\eta}{\sqrt{\rho^2+
\eta^2}}=0.903~,~cos(\theta_1-\theta_2)=\frac{-\rho}{\sqrt{\rho^2+\eta^2}}=
-0.430~,~\left|\frac{G_1}{G_2}\right|=(1-\frac{\lambda^2}{2})
\sqrt{\rho^2+\eta^2}$;
while in processes (17)-(30), $\sin(\theta_1-\theta_2)=\frac{-\eta}{\sqrt
{\rho^2+\eta^2}}~,~cos(\theta_1-\theta_2)=\frac{\rho}{\sqrt{\rho^2+\eta^2}}~,~
\left|\frac{G_1}{G_2}\right|=\frac{\lambda^2\sqrt{\rho^2+\eta^2}}{(1-\frac
{\lambda^2}{2})}$ which is doubly Cabibbo suppressed.  In processes (1)-(6) and
(17)-(20), $T_1$ is dominated by the tree diagram and $T_2$ only by the penguin
diagram only, so $|T_1|\gg|T_2|$ ; in processes (9)-(12) and (21)-(26),
$T_1$ and $T_2$ are both from the penguin diagram contribution only, so
$|T_1|
\sim |T_2|$ ; while in the processes (13)-(16) and (27)-(30), $T_2$ is
dominated by
the tree diagram and $T_1$ by the penguin diagram, so $|T_2|\gg|T_1|$.  Since
the CP violating asymmetry is proportional to $sin\zeta=\frac{2h}{1+h^2}$, where
$h=\left|\frac{G_1T_1}{G_2T_2}\right|$, if $h\gg 1$ or $h\ll 1$, $sin \zeta$
is small. From Table II, The CP violating asymmetry is suppressed
strongly by the small $sin\zeta$ in the processes $B_c^-\rightarrow \psi
D^{-}~,~\eta_{c}D_s^- ~,~\psi D_s^- ~,~\eta_{c}D_s^{*-}~,~\psi D_s^{*-}$ .  For
the processes with $h\sim O(1)~(~|G_1T_1|\sim |G_2T_2|~)$, $sin\zeta$ is large,
such as in the processes $B_c^-\rightarrow \pi^0 D^- ~,~\rho^0 D^-~,~ \rho^0
D^{*-}~,~ \bar{D}^{*0}K^-~,~\bar{D}^0 K^{*-}~,~\bar{D}^{*0}K^{*-}$ .  The hard
strong phase$(\delta_1-\delta_2)$ is caused by the interference between
penguin and tree diagrams or penguin diagram themselves, which rely on the
penguin loop-integral functions $G_{u}(k^2)$ and $G_{c}(k^2)$.  As discussed in
Section IV, we take the average value of the momentum squared 
carried by virtual particles for our calculation.  In most processes (below the
$c\bar{c}$ threshold) $\frac{\bar{k^2}}{m_b^2}\sim 0.5$ , but for producing
$c\bar{c}$ pair $(~B_c^-\rightarrow (c\bar{c})D^-~,~(c\bar{c})D^{*-}
~,~(c\bar{c})D_s^-~,~(c\bar{c})D_s^{*-}~),~\frac{\bar{k^2}}{m_b^2}\sim 0.7$ .
From Table.II, we can see that for the processes with interference of penguin
diagram themselves only, such as the processes (9)-(12) and (21)-(26), $|sin
(\delta_1-\delta_2)|$ is smaller than that for the processes with interference
between penguin and tree diagrams.  So, if the decay processes both satisfy:
(i) $|G_1T_1|\sim|G_2T_2|$ , (ii) with interference between penguin and tree
diagrams, then the CP violating asymmetry will be large, such as the processes
(5)-(8) and (17)-(20).

  According to our CP violating asymmetry formula ${\cal A}_{cp}=\frac{{\cal A}
_{0}}{1+{\cal A}_{1}}$, where ${\cal A}_{0}=-sin\zeta sin(\theta_1-\theta_2)
sin(\delta_1-\delta_2)~,~{\cal A}_{1}=sin\zeta cos(\theta_1-\theta_2)cos
(\delta_1-\delta_2)$ , it can be seen that ${\cal A}_{cp}$ is proportional to
${\cal A}_{0}$ and the proportion factor is $\frac{1}{1+{\cal A}_{1}}$.  If
$|{\cal A}_{1}|\ll 1$ , then ${\cal A}_{cp}\sim{\cal A}_{0}$ .  Since ${\cal A}
_{0}\propto\eta~,~{\cal A}_{1}\propto\rho$ , the CP violating asymmetry is more
sensitive to the parameter $\eta$ than the parameter $\rho$.  In this paper, we
have taken $\rho=0.160$ [5] for numerical estimate, the main difference compare
with the early literature is to change the sign of the parameter $\rho$.  For
the case of $|{\cal A}_{1}|\ll 1$ , the influence of this change is very small,
but in the case of $|{\cal A}_{1}|\sim 1$ , it is important.  In the processes
(5)-(8), the CP violating asymmetry will be smaller than that with minus $\rho$;
while in the processes (17)-(20), it will be larger than that with minus $\rho$.

  In this paper , we do not discuss the decay width of the $B_c$ meson, since it
needs to calculate the hadronic matrix elements which is strongly 
model-dependent. But for testing the CP violation in experiment, we need to find 
processes with both larger CP violating asymmetry and larger branching ratio.  
For three standard deviation $(3\sigma)$ signature $\varepsilon_{f}N\sim
\frac{9}{B_r{\cal A}_{cp}^2}$, where $\varepsilon_{f}$ is the detecting
efficiency of the final state, $N$ is number of the $B_c^{\pm}$ . Since lack of
experimental data for the branching fraction of $B_c$ decays at present, so 
in Table.II, we use the results of decay widths for the hadronic decay of the 
$B_c^-$ calculated in Ref.[4] but take $\tau_{B_c}$=0.5 ps [14] to estimate
the branching ratio $B_r$ and $\varepsilon_{f}N$. This estimation show that 
in experiment the best decay modes to observe CP violation in $B_c$ decays 
would be $B_c^-\rightarrow{\bar D}^{*0}K^{*-}~,~{\bar D}^{0}K^{*-}~,~{\bar D}^
{*0}K^{-}~,~{\bar D}^{0}K^-$ and $B_c^-\rightarrow\eta_{c}D^-$, which need 
about $10^8$ of $B_c^{\pm}$ events. 
 
  In summary, we have calculated the CP violating asymmetries for the $B_c$
meson decays into PP, PV, VV mesons.  It is only relevant to the three angles,
weak phase $(\theta_1-\theta_2)$, strong phase $(\delta_1-\delta_2)$ and
the angle $\zeta$.  If  $|sin (\delta_1-\delta_2)|$ and $sin\zeta$ 
are both large ($|sin(\theta_1-\theta_2)|$ is the same for all decay
processes ), the CP
violating asymmetry would be large.  There are less uncertainties in our
calculated CP
violating asymmetries since we avoided calculating the hadronic matrix
elements in our method. In general, we need
about
$10^8$ of $B_c^{\pm}$ events for testing the CP violation in experiment.

  Finally we should mention that, in this paper, we only calculated the hard
strong phases, the effects of the soft strong phases would also be important, it
needs further study.

\section* {Acknowledgements}
\noindent

  This work was supported in part by National Natural Science Foundation of China
and Beijing Electron Positron Collider National Laboratory.

\newpage \section*{Table Captions} 
\noindent

Table.I. The quark-diagram amplitudes for the $B_c^-$ decays into PP, PV, VV
	 mesons in the spectator approximation.\\

Table.II. The CP violating asymmetries for two-body mesonic decays of the
	  $B_c^-$ meson.  where $h=\left |\frac{G_1T_1}{G_2T_2}\right
	  |,~sin\zeta=\frac{2h} {1+h^2};~{\cal A}_{cp}=\frac{{\cal A}_0}{1+{\cal
	  A}_1},~{\cal A}_0 =-sin\zeta sin(\theta_1-\theta_2)
	  sin(\delta_1-\delta_2),~{\cal A}_1= sin\zeta cos(\theta_1-\theta_2)
	  cos(\delta_1-\delta_2)$. In processes (1)-(16), $sin(\theta_1-
	  \theta_2)$=0.903 , $cos(\theta_1-\theta_2)$=-0.430 ; while in 
	  processes (17)-(30), $sin(\theta_1-\theta_2)$=-0.903 , $cos(\theta_1-
	  \theta_2)$=0.430 .
	   
\newpage

\begin{center} Table.I \\ 
\vspace{0.5cm} 
\begin{scriptsize}

\begin{tabular}{ccc}
\hline No.  & Final state $f$ & The quark-diagram amplitudes\\ \hline 
(1) & $\bar{D}^0\pi^-$ & ~ \\ 
(2) & $\bar{D}^{*0}\pi^-$ & $V_{ud}^*V_{ub}[~T(a^{tree})+T(a_u^{pen})~]+ 
      V_{cd}^*V_{cb}T(a_c^{pen})$\\ 
(3) & $\bar{D}^0\rho^-$ & ~ \\ 
(4) & $\bar{D}^{*0}\rho^-$ & ~ \\ \hline 
(5) & $\pi^0D^-$ & ~ \\ 
(6) & $\pi^0D^{*-}$ & $V_{ud}^*V_{ub}[~T(b^{tree})+T(b_u^{pen})~]+ 
      V_{cd}^*V_{cb}T(b_c^{pen})$\\ 
(7) &$\rho^0D^-$ & ~ \\ 
(8) & $\rho^0D^{*-}$ & ~ \\ \hline 
(9) & $K^0D_s^-$ & ~ \\
(10)& $K^0D_s^{*-}$ & $V_{ud}^*V_{ub}T(b_u^{pen})+V_{cd}^*V_{cb}T(b_c^{pen})$\\
(11)& $K^{*0}D_s^-$ & ~ \\ 
(12)& $K^{*0}D_s^{*-}$ & ~\\ \hline 
(13)& $\eta_cD^-$ & ~ \\ 
(14)& $\psi D^-$ & $V_{ud}^*V_{ub}T(a_u^{pen})+V_{cd}^*V_{cb}[~T(a^{tree})
      +T(b^{tree})+T(a_c^{pen})~]$\\ 
(15)& $\eta_cD^{*-}$ & ~ \\ 
(16)& $\psi D^{*-}$ & ~ \\ \hline 
(17)& $\bar{D}^0K^-$ & ~ \\ 
(18)& $\bar{D}^{*0}K^-$ & $V_{us}^*V_{ub}[~T(a^{tree})+T(a_u^{pen})~]+
      V_{cs}^*V_{cb}T(a_c^{pen})$\\ 
(19)& $\bar{D}^0K^{*-}$ & ~ \\ 
(20)& $\bar{D}^{*0}K^{*-}$ & ~ \\ \hline 
(21)& $\bar{K}^0D^-$ & ~ \\ 
(22)& $\bar{K}^0D^{*-}$ & ~ \\ 
(23)& $\bar{K}^{*0}D^-$ & $V_{us}^*V_{ub}T(b_u^{pen})+
      V_{cs}^*V_{cb}T(b_c^{pen})$\\ 
(24)& $\bar{K}^{*0}D^{*-}$ & ~ \\ 
(25)& $\phi D_s^-$ & ~ \\ 
(26)& $\phi D_s^{*-}$ & ~ \\ \hline 
(27)& $\eta_cD_S^-$ & ~ \\
(28)& $\psi D_s^-$ & $V_{us}^*V_{ub}T(a_u^{pen})+
      V_{cs}^*V_{cb}[~T(a^{tree})+T(b^{tree})+T(a_c^{pen})~]$\\ 
(29)& $\eta_cD_s^{*-}$ & ~ \\ 
(30)& $\psi D_s^{*-}$ & ~ \\ \hline 
\end{tabular} 
\end{scriptsize}
\end{center}

\newpage 
\begin{center} 
Table.II\\ 
\vspace{0.5cm} 
\begin{scriptsize} 
\tabcolsep 0.04 in 
\begin{tabular}{ccccccccccccc} \hline 
No.&$f$ &$\xi_f$ &$\frac{\bar{k^2}}{m_b^2}$ &$sin(\delta_1-\delta_2)$ &$h$ 
   &$sin\zeta$ &${\cal A}_0$ &${\cal A}_1$ &${\cal A}_{cp}=\frac{{\cal A}_0}
   {1+{\cal A}_1}$ &~~~~&$B_r$ &$\varepsilon_{f}N$ \\ \hline 
(1) &$\bar{D}^0\pi^-$ &0.545 &0.498 &-0.197 &5.284 &0.365 &0.0649 &0.154 
    &$5.62\times 10^{-2}$& &$2.68\times 10^{-6}$ &$1.06\times 10^9$\\
(2) &$\bar{D}^{*0}\pi^-$ &-0.544 &0.498 &-0.274 &27.93 &0.0715 &0.0177 &0.0296 
    &$1.72\times 10^{-2}$ & &$2.18\times 10^{-6}$ &$1.40\times 10^{10}$\\ 
(3) &$\bar{D}^0\rho^-$ &0 &0.487 &-0.202 &9.010 &0.219 &0.0399 &0.0922 
    &$3.65\times 10^{-2}$ & &$6.85\times 10^{-6}$ &$9.86\times 10^8$\\ 
(4) &$\bar{D}^{*0}\rho^-$ &0 &0.487 &-0.202 &9.010 &0.219 &0.0399 &0.0922 
    &$3.65\times 10^{-2}$ & &$6.81\times 10^{-6}$ &$9.92\times 10^8$\\ \hline 
(5) &$\pi^0D^-$ &0.380 &0.500 &-0.257 &1.462 &0.932 &0.216 &0.387 
    &$1.56\times 10^{-1}$ & &$2.28\times 10{-7}$ &$1.62\times 10^9$\\ 
(6) &$\pi^0D^{*-}$ &-0.379 &0.500 &-0.316 &5.528 &0.350 &0.0999 &0.143 
    &$8.74\times 10^{-2}$ & &$7.17\times 10^{-8}$ &$1.64\times 10^{10}$\\ 
(7) &$\rho^0D^-$ &0 &0.487 &-0.255 &2.401 &0.710 &0.163 &0.295 
    &$1.26\times 10^{-1}$ & &$4.21\times 10^{-7}$ &$1.35\times 10^9$\\ 
(8) &$\rho^0D^{*-}$ &0 &0.487 &-0.255 &2.401 &0.710 &0.163 &0.295
    &$1.26\times 10^{-1}$ & &$1.39\times 10^{-6}$ &$4.08\times 10^8$\\ \hline 
(9) &$K^0D_s^-$ &0.577 &0.557 &-0.0616 &0.3227 &0.585 &0.0325 &-0.251 
    &$4.34\times 10^{-2}$ & &$2.38\times 10^{-7}$ &$2.01\times 10^{10}$\\
(10)&$K^0D_s^{*-}$ &-0.537 &0.557 &-0.189 &0.3066 &0.561 &0.0957 &-0.237 
    &$1.25\times 10^{-1}$& &$2.67\times 10^{-8}$ &$2.16\times 10^{10}$\\ 
(11)&$K^{*0}D_s^-$ &0 &0.503 &-0.0892 &0.3161 &0.575 &0.0463 &-0.246 
    &$6.14\times 10^{-2}$ & &$1.43\times 10^{-7}$ &$1.67\times 10^{10}$\\
(12)&$K^{*0}D_s^{*-}$ &0 &0.503 &-0.0892 &0.3161 &0.575 &0.0463 &-0.246 
    &$6.14\times 10^{-2}$ & &$2.85\times 10^{-7}$ &$8.38\times 10^9$\\ \hline 
(13)&$\eta_cD^-$ &1.489 &0.744 &0.261 & 0.03055 &0.0610 &-0.0144&0.0253 
    &$1.40\times 10^{-2}$ & &$2.95\times 10^{-4}$ &$1.56\times 10^8$\\
(14)&$\psi D^-$ &-0.835 &0.744 &0.579 &0.0007797 &0.00156 &-0.000816 &-0.000547
    &$8.16\times 10^{-4}$ & &$2.19\times 10^{-5}$ &$6.17\times 10^{11}$\\ 
(15)&$\eta_cD^{*-}$ &0 &0.699 &0.286 &0.01012 &0.0202 &-0.00522 &0.00832 
    &$5.18\times 10^{-3}$ & &$6.89\times 10^{-5}$ &$4.87\times 10^9$\\ 
(16)&$\psi D^{*-}$ &0 &0.699 &0.286 &0.01012 &0.0202 &-0.00522 &0.00832 
    &$5.18\times 10^{-3}$ & &$4.95\times 10^{-4}$ &$6.78\times 10^8$\\ \hline
(17)&$\bar{D}^0K^-$ &0.565 &0.433 &-0.173 &0.2579 &0.484 &-0.0756 &-0.205 
    &$-9.51\times 10^{-2}$ & &$4.67\times 10^{-6}$ &$2.13\times 10^8$\\ 
(18)&$\bar{D}^{*0}K^-$&-0.564 &0.433 &-0.243 &1.482 &0.927 &-0.203 &-0.387
    &$-3.31\times 10^{-1}$ & &$4.89\times 10^{-7}$ &$1.68\times 10^8$\\ 
(19)&$\bar{D}^0K^{*-}$ &0 &0.464 &-0.194 &0.4537 &0.753 &-0.132 &-0.318 
    &$-1.94\times 10^{-1}$ & &$3.58\times 10^{-6}$ &$6.68\times 10^7$\\
(20)&$\bar{D}^{*-}K^{*-}$ &0 &0.464 &-0.194 &0.4537 &0.753 &-0.132 &-0.318 
    &$-1.94\times 10^{-1}$ & &$3.87\times 10^{-6}$ &$6.18\times 10^7$\\ \hline 
(21)&$\bar{K}^0D^-$ &0.557&0.434 &-0.0975 &0.01601 &0.0320 &-0.00282 &0.0137 
    &$-2.78\times10^{-3}$ & &$5.39\times 10^{-6}$ &$2.16\times 10^{11}$\\ 
(22)&$\bar{K}^0D^{*-}$ &-0.555 &0.434 &-0.288 &0.01504 &0.0301 &-0.00783 &0.0124
    &$-7.73\times 10^{-3}$ & &$3.05\times 10^{-7}$ &$4.94\times 10^{11}$\\
(23)&$\bar{K}^{*0}D^-$ &0 &0.464 &-0.103 &0.01593 &0.0319&-0.00297 &0.0136 
    &$-2.93\times 10^{-3}$ & &$3.14\times 10^{-6}$ &$3.34\times 10^{11}$\\ 
(24)&$\bar{K}^{*0}D^{*-}$ &0&0.464 &-0.103  &0.01593 &0.0319 &-0.00297 &0.0136 
    &$-2.93\times 10^{-3}$ & &$3.34\times 10^{-6}$ &$3.14\times 10^{11}$\\ 
(25)&$\phi D_s^-$ &0 &0.480 &-0.0967 &0.01599&0.0320 &-0.00279 &0.0137 
    &$-2.75\times 10^{-3}$ & &$1.64\times 10^{-6}$ &$7.26\times 10^{11}$\\ 
(26)&$\phi D_s^{*-}$ &0&0.480 &-0.0967 &0.01599 &0.0320 &-0.00279 &0.0137
    &$-2.75\times 10^{-3}$ & &$4.23\times 10^{-6}$ &$2.81\times 10^{11}$\\ \hline 
(27)&$\eta_cD_s^-$ &1.474 &0.708 &0.260 &0.001549 &0.00310 &0.000728 &-0.00129 
    &$7.27\times 10^{-4}$ & &$7.27\times 10^{-3}$ &$2.34\times 10^9$\\ 
(28)&$\psi D_s^-$ &-0.827 &0.708 &0.664 &0.00003624 &0.0000725 &0.0000435 
    &0.0000233 &$4.35\times 10^{-5}$ & &$6.48\times 10^{-4}$ &$7.34\times 
    10^{12}$\\
(29)&$\eta_cD_s^{*-}$ &0 &0.668 &0.285 &0.0005167 &0.00103 &0.000265 &-0.000425
    &$2.65\times 10^{-4}$ & &$1.67\times 10^{-3}$ &$7.67\times 10^{10}$\\ 
(30)&$\psi D_s^{*-}$ &0 &0.668 &0.285 &0.0005167 &0.00103 &0.000265 &-0.000425 
    &$2.65\times 10^{-4}$ & &$1.33\times 10^{-2}$ &$9.64\times 10^9$\\ \hline 
\end{tabular} 
\end{scriptsize} 
\end{center}

\newpage 
\section*{Figure captions} 
\noindent

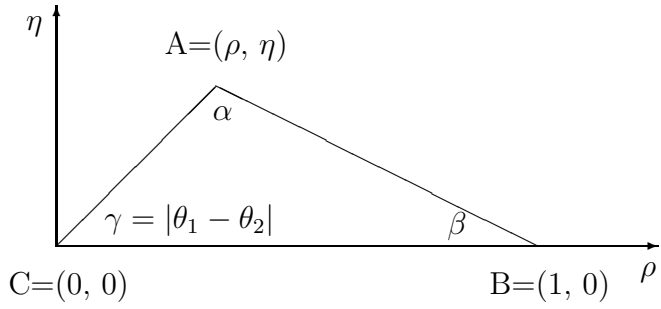
Fig. 1. The unitarity triangle of the CKM matrix elements
	$V_{ub}^*V_{ud}+V_{cb}^*V_{cd}+V_{tb}^*V_{td}=0$ in the ($\rho$ ,
	$\eta$) plane.\\

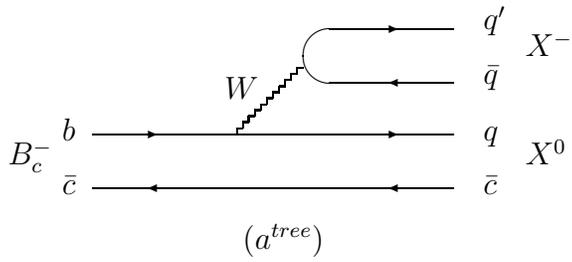
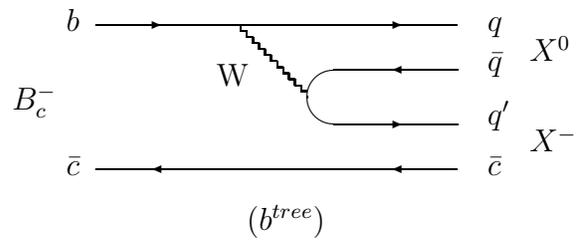
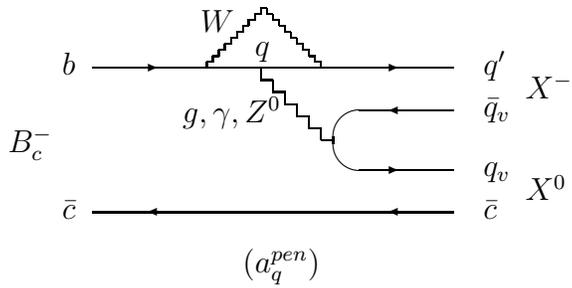
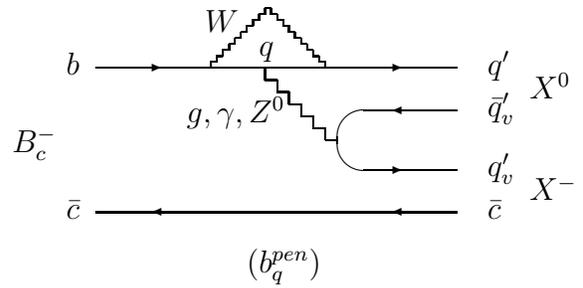
Fig. 2. Quark diagrams for $B_c^-$ decaying into two mesons $X^0$ and $X^-$ in
	the spectator approximation.  $a^{tree}$ is the color-favored tree
	diagram; $b^{tree}$ is the color-suppressed tree diagram; $a_q^{pen}$
	and $b_q^{pen}$ are the timelike penguin diagrams.  Where $q,q_v=u,c$ ;
	$q',q'_v=d,s$, the subscripts "$v$" denote "vacuum".\\

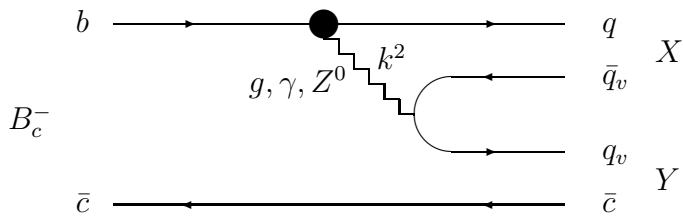
Fig. 3. The timelike penguin diagrams for $B_c^-$ decaying into two mesons $X$
	and $Y$.  The dark dot denotes the W-loop, subscript "$v$" denote
	"vacuum".  $k^2$ is the momentum squared carried by the virtual gluon,
	photon and $Z^0$.

\newpage \setlength{\unitlength}{0.08 mm} \begin{picture}(1500,400)
\put(0,0){\vector(1,0){1000}} \put(0,0){\vector(0,1){400}}
\put(0,0){\line(1,1){265}} \put(800,0){\line(-2,1){535}} \put(-80,-80){C=(0, 0)}
\put(720,-80){B=(1, 0)} \put(180,320){A=($\rho$, $\eta$)} \put(-50,360){$\eta$}
\put(970,-50){$\rho$} \put(80,30){$\gamma=|\theta_1-\theta_2|$}
\put(650,20){$\beta$} \put(260,210){$\alpha$} \put(400,-160){Fig.  1.}
\end{picture}

\setlength{\unitlength}{0.08 mm} \begin{picture}(1500,1500)
\put(0,0){\line(1,0){600}} \put(0,240){\line(1,0){600}}
\multiput(280,330)(-10,-10){10}{\line(0,1){10}}
\multiput(290,330)(10,-10){10}{\line(0,1){10}}
\multiput(280,330)(-10,-10){9}{\line(-1,0){10}} \multiput(280,340)(10,
-10){10}{\line(1,0){10}} \put(600,120){\oval(400,100)[l]}
\multiput(280,240)(20,-20){6}{\line(0,-1){20}}
\multiput(280,240)(20,-20){7}{\line(-1,0){20}} \put(100,0){\vector(-1,0){10}}
\put(500,0){\vector(-1,0){10}} \put(100,240){\vector(1,0){10}}
\put(500,240){\vector(1,0){10}} \put(500,170){\vector(-1,0){10}}
\put(500,70){\vector(1,0){10}} \put(150,150){$g,\gamma,Z^0$} \put(270,260){$q$}
\put(180,300){$W$} \put(-50,-10){$\bar{c}$} \put(-50,230){$b$}
\put(-140,100){$B_c^-$} \put(650,-10){$\bar{c}$} \put(650,230){$q'$}
\put(650,160){$\bar{q}_v$} \put(650,60){$q_v$} \put(720,190){$X^-$}
\put(720,20){$X^0$} \put(250,-100){$(a_q^{pen})$}

\put(1200,0){\line(1,0){600}} \put(1200,240){\line(1,0){600}}
\multiput(1480,330)(-10,-10){10}{\line(0,1){10}}
\multiput(1490,330)(10,-10){10}{\line(0,1){10}}
\multiput(1480,330)(-10,-10){9}{\line(-1,0){10}}
\multiput(1480,340)(10,-10){10}{\line(1,0){10}}
\put(1800,120){\oval(400,100)[l]}
\multiput(1480,240)(20,-20){6}{\line(0,-1){20}}
\multiput(1480,240)(20,-20){7}{\line(-1,0){20}} \put(1300,0){\vector(-1,0){10}}
\put(1700,0){\vector(-1,0){10}} \put(1300,240){\vector(1,0){10}}
\put(1700,240){\vector(1,0){10}} \put(1700,170){\vector(-1,0){10}}
\put(1700,70){\vector(1,0){10}} \put(1350,150){$g,\gamma,Z^0$}
\put(1470,260){$q$} \put(1380,300){$W$} \put(1150,-10){$\bar{c}$}
\put(1150,230){$b$} \put(1060,100){$B_c^-$} \put(1850,-10){$\bar{c}$}
\put(1850,230){$q'$} \put(1850,160){$\bar{q}'_v$} \put(1850,60){$q'_v$}
\put(1920,190){$X^0$} \put(1920,20){$X^-$} \put(1450,-100){$(b_q^{pen})$}

\put(0,700){\line(1,0){600}} \put(0,790){\line(1,0){600}}
\multiput(240,790)(10,10){11}{\line(0,1){10}}
\multiput(240,790)(10,10){12}{\line(-1,0){10}} \put(600,920){\oval(500,90)[l]}
\put(100,700){\vector(-1,0){10}} \put(500,700){\vector(-1,0){10}}
\put(100,790){\vector(1,0){10}} \put(500,790){\vector(1,0){10}}
\put(500,875){\vector(-1,0){10}} \put(500,965){\vector(1,0){10}}
\put(220,850){$W$} \put(-50,690){$\bar{c}$} \put(-50,780){$b$}
\put(-140,740){$B_c^-$} \put(650,690){$\bar{c}$} \put(650,780){$q$}
\put(650,875){$\bar{q}$} \put(650,965){$q'$} \put(720,920){$X^-$}
\put(720,740){$X^0$} \put(250,600){$(a^{tree})$}

\put(1200,700){\line(1,0){600}} \put(1200,940){\line(1,0){600}}
\multiput(1440,940)(10,-10){11}{\line(0,-1){10}}
\multiput(1440,940)(10,-10){12}{\line(-1,0){10}}
\put(1800,820){\oval(500,90)[l]} \put(1300,700){\vector(-1,0){10}}
\put(1700,700){\vector(-1,0){10}} \put(1300,940){\vector(1,0){10}}
\put(1700,940){\vector(1,0){10}} \put(1700,865){\vector(-1,0){10}}
\put(1700,775){\vector(1,0){10}} \put(1400,840){W} \put(1150,690){$\bar{c}$}
\put(1150,930){$b$} \put(1060,800){$B_c^-$} \put(1850,690){$\bar{c}$}
\put(1850,930){$q$} \put(1850,865){$\bar{q}$} \put(1850,775){$q'$}
\put(1920,880){$X^0$} \put(1920,730){$X^-$} \put(1450,600){$(b^{tree})$}
\put(850,-220){Fig.  2.}  \end{picture}

\setlength{\unitlength}{0.1 mm} \begin{picture}(1500,600)
\put(0,0){\line(1,0){600}} \put(0,240){\line(1,0){600}}
\put(600,120){\oval(400,100)[l]} \multiput(280,240)(20,-20){6}{\line(0,-1){20}}
\multiput(280,240)(20,-20){7}{\line(-1,0){20}} \put(280,240){\circle*{40}}
\put(100,0){\vector(-1,0){10}} \put(500,0){\vector(-1,0){10}}
\put(100,240){\vector(1,0){10}} \put(500,240){\vector(1,0){10}}
\put(500,170){\vector(-1,0){10}} \put(500,70){\vector(1,0){10}}
\put(180,150){$g,\gamma,Z^0$} \put(350,180){$k^2$} \put(-50,-10){$\bar{c}$}
\put(-50,230){$b$} \put(-140,100){$B_c^-$} \put(650,-10){$\bar{c}$}
\put(650,230){$q$} \put(650,160){$\bar{q}_v$} \put(650,60){$q_v$}
\put(720,190){$X$} \put(720,20){$Y$} \put(280,-100){Fig.  3.}  \end{picture}

\end{document}